\documentclass[10pt,preprint]{emulateapj}
\usepackage{url}
\usepackage{color}
\usepackage{natbib}
\newcommand{\fig}[1]{Figure~\ref{#1}}

\newcommand{\speed}[1]{#1 km~s${}^{-1}$}

\newcommand{\acc}[1]{#1 m~s${}^{-2}$}

\begin{document}

\shorttitle{} %

\shortauthors{Shen et al.}

\title{Diffraction, Refraction, and Reflection of An Extreme-Ultraviolet Wave Observed during Its Interactions with Remote Active Regions}

\author{Yuandeng Shen\altaffilmark{1,2,3,4}, Yu Liu\altaffilmark{1,3,4}, Jiangtao Su\altaffilmark{3}, Hui Li\altaffilmark{4}, Ruijuan Zhao\altaffilmark{1}, Zhanjun Tian\altaffilmark{1}, Kiyoshi Ichimoto\altaffilmark{2}, and Kazunari Shibata\altaffilmark{2}}

\altaffiltext{1}{Yunnan Astronomical Observatory, Chinese Academy of Sciences, Kunming 650011, China; ydshen@ynao.ac.cn}
\altaffiltext{2}{Kwasan and Hida Observatories, Kyoto University, Kyoto 6078471, Japan}
\altaffiltext{3}{Key Laboratory of Solar Activity, Chinese Academy of Sciences, Beijing 100012, China}
\altaffiltext{4}{Key Laboratory of Dark Matter and Space Astronomy, Chinese Academy of Sciences, Nanjing 210008, China}
\slugcomment{Accepted by APJL, July 23, 2013}

\begin{abstract}
We present observations of the diffraction, refraction, and reflection of a global extreme-ultraviolet (EUV) wave propagating in the solar corona. These intriguing phenomena are observed when the wave interacts with two remote active regions, and they together exhibit the wave property of this EUV wave. When the wave approached AR11465, it became weaker and finally disappeared in the active region, but a few minutes latter a new wavefront appeared behind the active region, and it was not concentric with the incoming wave. In addition, a reflected wave was also observed simultaneously on the wave incoming side. When the wave approached AR11459, it transmitted through the active region directly and without reflection. The formation of the new wavefront and the transmission could be explained with diffraction and refraction effects, respectively. We propose that the different behaviors observed during the interactions may caused by different speed gradients at the boundaries of the two active regions. For the origin of the EUV wave, we find that it formed ahead of a group of expanding loops a few minutes after the start of the loops' expansion, which represents the initiation of the associated coronal mass ejection (CME). Based on these results, we conclude that the EUV wave should be a nonlinear magnetosonic wave or shock driven by the associated CME, which propagated faster than the ambient fast-mode speed and gradually slowed down to an ordinary linear wave. Our observations support the hybrid model that includes both fast wave and slow non-wave components.
\end{abstract}


\section{INTRODUCTION}
In recent years, a hot debating topic in solar physics is the global extreme-ultraviolet (EUV) wave, which is propagating broad and diffuse bright structure in the solar corona. Since the typical speed \citep[\speed{200 -- 400};][]{thom09} is usually higher than the quiet-Sun sound speed, the EUV wave was interpreted as fast magnetosonic wave at first \citep{thom98,thom99}. On the other hand, it was proposed as the expecting coronal counterpart of the chromospheric Moreton wave, which had been explained as the intersection between a fast-mode coronal wave and the chromosphere \citep{uchi68}. For years, solar physicists are debating about the real physical nature (i.e., wave or non-wave) and origin (i.e., flare or CME) of the EUV wave. However, these basic but important questions are still hanging in the air.

The fast wave scenario has been supported by a number of observational \citep{vero10,koza11,asai12,liu12,shen12a,chen12,zhen11,li12,olme12,kwon13} and numerical studies \citep{wang00,wu01,ofma02,mei12}. However, this is challenged by a few non-wave explanations \citep{dela00,harr03,attr09,schr11,dela08,attr07}. Furthermore, In some cases, both a slow and a fast wave components could be observed simultaneously \citep{zhuk04,chen11,shen12b}, and they could be interpreted with the so-called hybrid model \citep{chen02,chen05,cohe09,schm10,down11}. The classification of EUV waves could be made according to their kinematical behavior, i.e., fast, moderate, and slow waves \citep{warm11}. The first two classes could be interpreted using wave models, while the last one often shows non-wave characteristics. Detailed observational characteristics and theoretical interpretations of EUV wave could be found in recent reviews \citep{warm10,gall11,pats12}.

Studying the interaction between EUV waves and other magnetic structures is an effective way to identify the real nature of EUV waves. \cite{gopa09} reported an EUV wave that was reflected by a remote coronal hole. \cite{liu10} observed an EUV wave that has both slow and fast components, and multiple ripples are produced when the fast component overtakes the slow one. In another case, they evidenced that an EUV wave can penetrate into cavity structure and result in the oscillation of an imbed filament\citep{liu12}. The launching of filament and loop oscillations by an EUV wave were also observed \citep[e.g.,][]{asai12,shen12a}. With stereoscopic observations, \cite{olme12} reported the reflection from and transmission through a coronal hole of an EUV wave. All these studies confirm the fast wave scenario. However, a few non-wave effects were also observed in some similar studies \citep[e.g.,][]{vero06}. To distinguish the real nature of EUV waves, more observational studies using high temporal and spatial resolution observations are necessary.

In this letter, we present an EUV wave on 2012 April 23, which was captured by the Atmospheric Imaging Assembly \citep[AIA;][]{leme12} and the Helioseismic and Magnetic Imager \citep[HMI;][]{scho12} on board  {\em Solar Dynamics Observatory} \citep[{\em SDO};][]{pesn12}. The EUV wave firstly appeared at the southeastern periphery of AR11461 (N12, W20) in the northern hemisphere, then it interacted with AR11465 (S18, W00) and AR11459 (S15, W39) in the southern hemisphere. During the interactions, we find several arguments including diffraction, reflection and refraction effects supporting the fast wave scenario.

\section{RESULTS}
The EUV wave was accompanied by a {\em GOES} C2.0 flare and a halo CME. The start, peak, and end times of the flare are 17:37, 17:51, and 18:05 UT, respectively. According to the CDAW catalog\footnote{http://cdaw.gsfc.nasa.gov/CME\_list/index.html}, the average CME speed is about \speed{528}, while the angular width and position angle are 360${}^\circ$ and 235${}^\circ$ respectively. In addition, the EUV wave was followed by a quasi-periodic fast magnetosonic wave \citep[e.g.,][]{liu11,shen12c}. Here we confine our attention on the EUV wave and its interaction with AR11465 and AR11459.

\begin{figure*}[thbp]
\epsscale{0.8}
\plotone{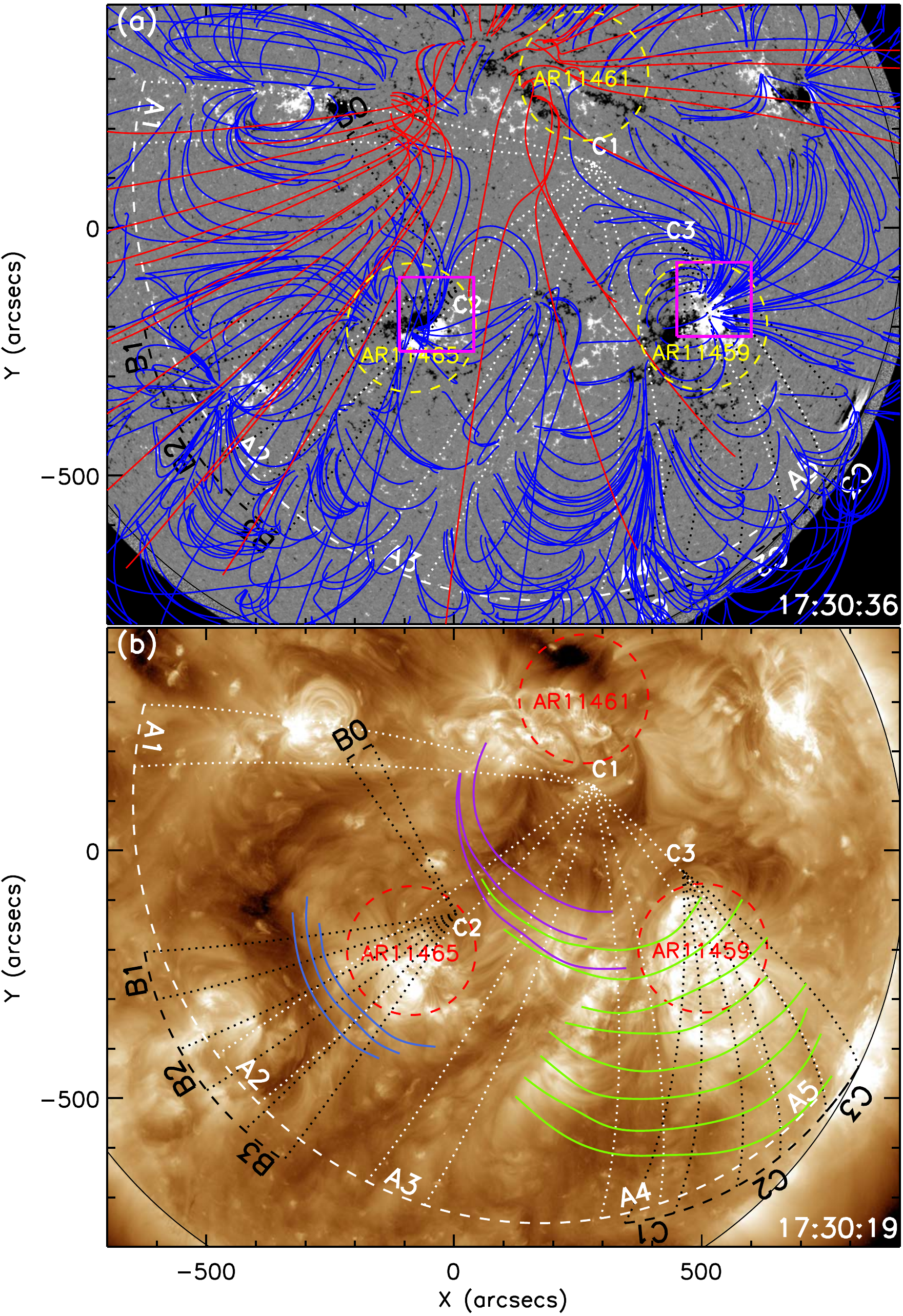}
\caption[]{\footnotesize
Top panel is an HMI LOS magnetogram overlaid with field lines extrapolated from the potential field source surface (PFSS) model \citep{schr03}, in which the blue (red) lines represent the closed (open) fields. Bottom panel is a pre-event 193 \AA\ image overlaid with wavefronts measured from 193 (magenta and blue) and 171 (green) \AA\ observations. Sectors A1 -- A5, B0, B1 -- B3, and C1 -- C3 are used to study the primary, reflected, diffracted, and refracted waves, respectively. The associated active regions are highlighted with dashed circles. 
\label{fig1}}
\end{figure*}

\fig{fig1} is an overview of the coronal condition in which the wave propagates. Several active regions can be identified on 2012 April 23, among which we are interested in AR11461, AR11465, and AR11459. The extrapolated coronal fields indicate that the north and east of AR11461 are dominated by open fields, while other regions are primarily closed fields (see \fig{fig1}(a)). The EUV wave first appeared at 17:42 UT on the southeast of AR11461 that produced the C2.0 flare. During the propagation, the EUV wave interacted with both AR11465 and AR11459. During the interaction with AR11465, secondary waves were observed simultaneously on the north and southeast of the active region (see the animation available in the online version of the journal). During the interaction with AR11459, transmission of the EUV wave through the active region could be identified. In addition, the northeast section of the wave front was obvious in 193 \AA\ observations, while the southwest segment was pronounced in 171 \AA\ images. Due to the hydrostatic weighting bias in the solar corona, higher layers are dominated by emission from hotter plasma. Since the 193 \AA\ channel shows hotter plasma than the 171 \AA\ channel, this phenomenon suggest that the propagation of the EUV wave was probably inclined to the solar surface.

\begin{figure*}[thbp]
\epsscale{0.89}
\plotone{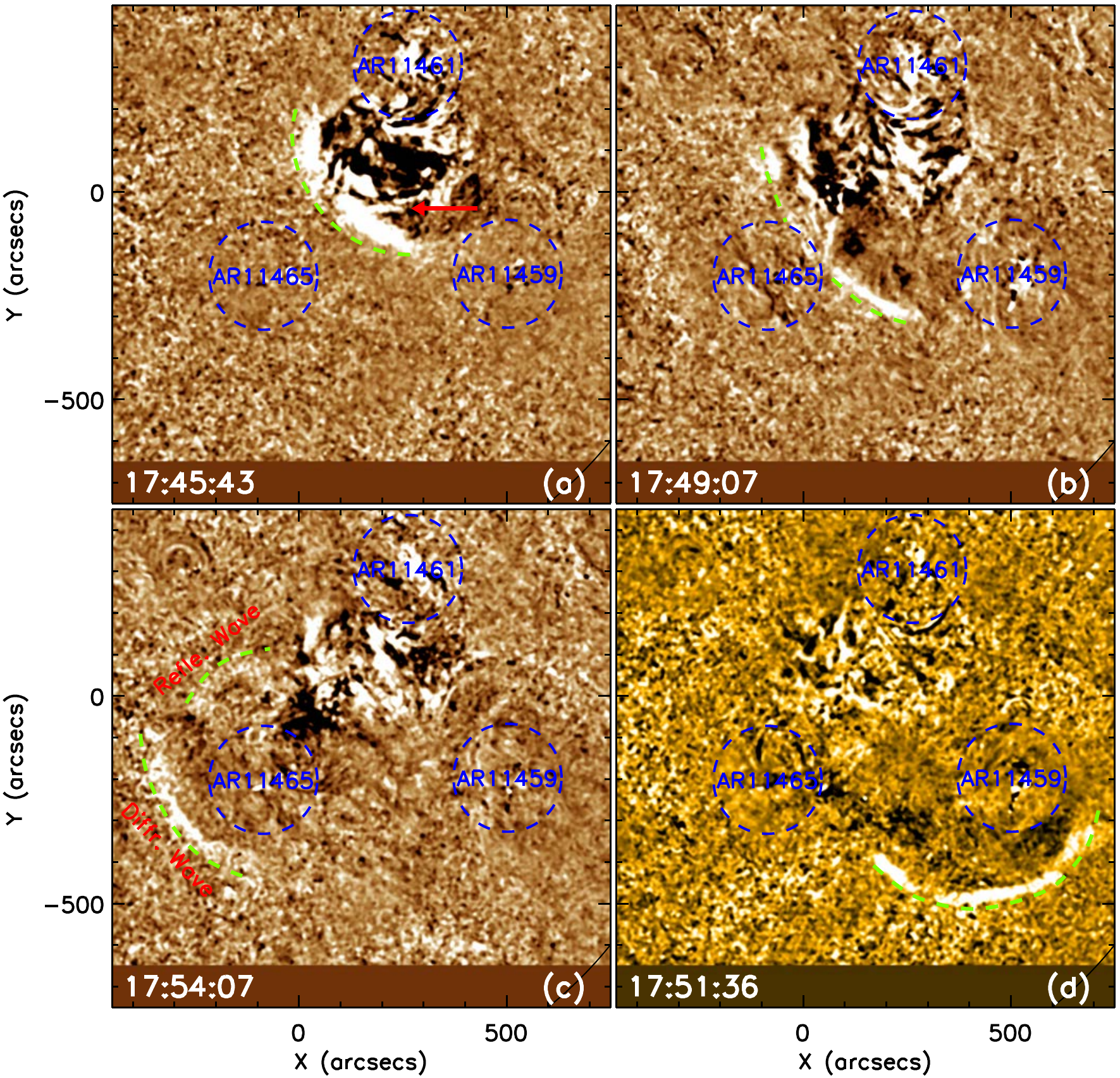}
\caption[]{\footnotesize
193 \AA\ (a) -- (c) and 171 \AA\ (d) running ratio images show the morphology evolution of the EUV wave, and the wavefronts are highlighted with dashed green curves. The red arrow in panel (a) indicates the expanding loops, while the three relevant active regions are indicated with dashed circles. An animation of this figure is available in the online journal.
\label{fig2}}
\end{figure*}

The morphology evolution of the wave is displayed in \fig{fig2}. At 17:45 UT, the wave had developed into a circular bright structure, behind which is an expanding loop system running ahead of a dimming region (see the red arrow in \fig{fig2} (a)). Here the expanding loops could be considered as the disk observation of the associated CME, which may excite the EUV wave ahead as identified in previous observations \citep[e.g.,][]{vero10,ma11,chen12}. The interaction of the wave with AR11465 occurred around 17:49 UT. As the wave approached AR11465, the wavefront became weaker and finally disappeared in the active region. A few minutes later, however, a new wavefront appeared on the southeast of AR11465, which is not concentric with the incoming EUV wave (see \fig{fig2} (c)). Since there was no other erupting source around that region, we propose that the new wavefront should be the diffracted wave generated during the interaction. In the meantime, reflected wave was also observed at the north of AR11465 (see \fig{fig2} (c)). When the EUV wave hits on AR11459, it directly transmitted through the active region and no reflected wave could be detected. The different behaviors of the EUV wave suggest the different physical conditions of the two active regions.

\begin{figure*}[thbp]
\epsscale{0.9}
\plotone{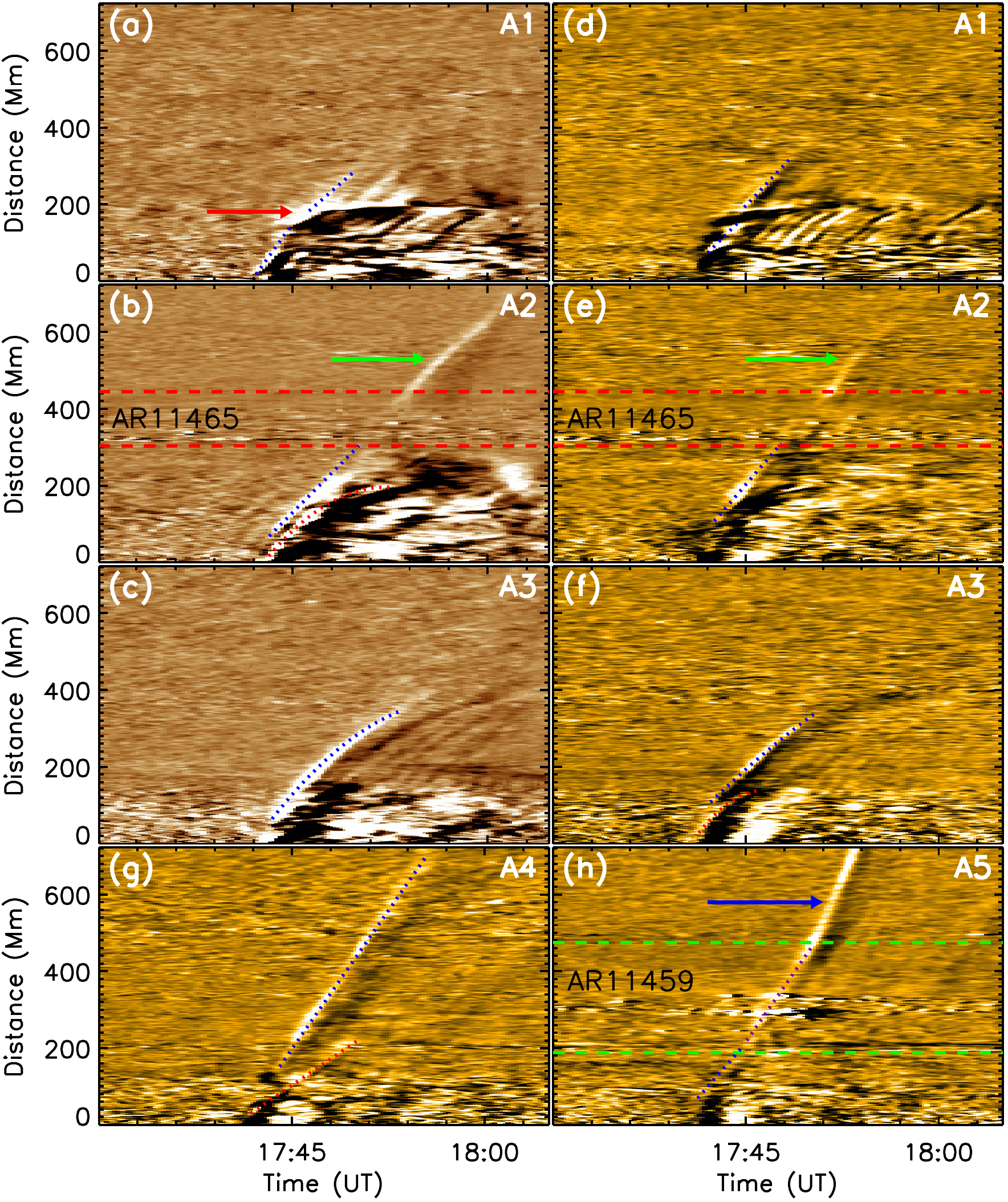}
\caption[]{\footnotesize
Time-distance plots along sectors A1 -- A5, in which (a) -- (c) and (d) -- (h) are obtained from 193 and 171 \AA\ running ratio images, respectively. The red (green) dashed lines indicate the boundaries of  AR11465 (AR11459). The dotted blue (red) lines are linear or quadratic fit to the EUV wave (expanding loops) ridges before the interactions with active regions. The red arrow points to the stationary wavefront, while the green (blue) arrow indicates the diffracted (refracted) wave.
\label{fig3}}
\end{figure*}
\begin{figure*}[thbp]
\epsscale{0.9}
\plotone{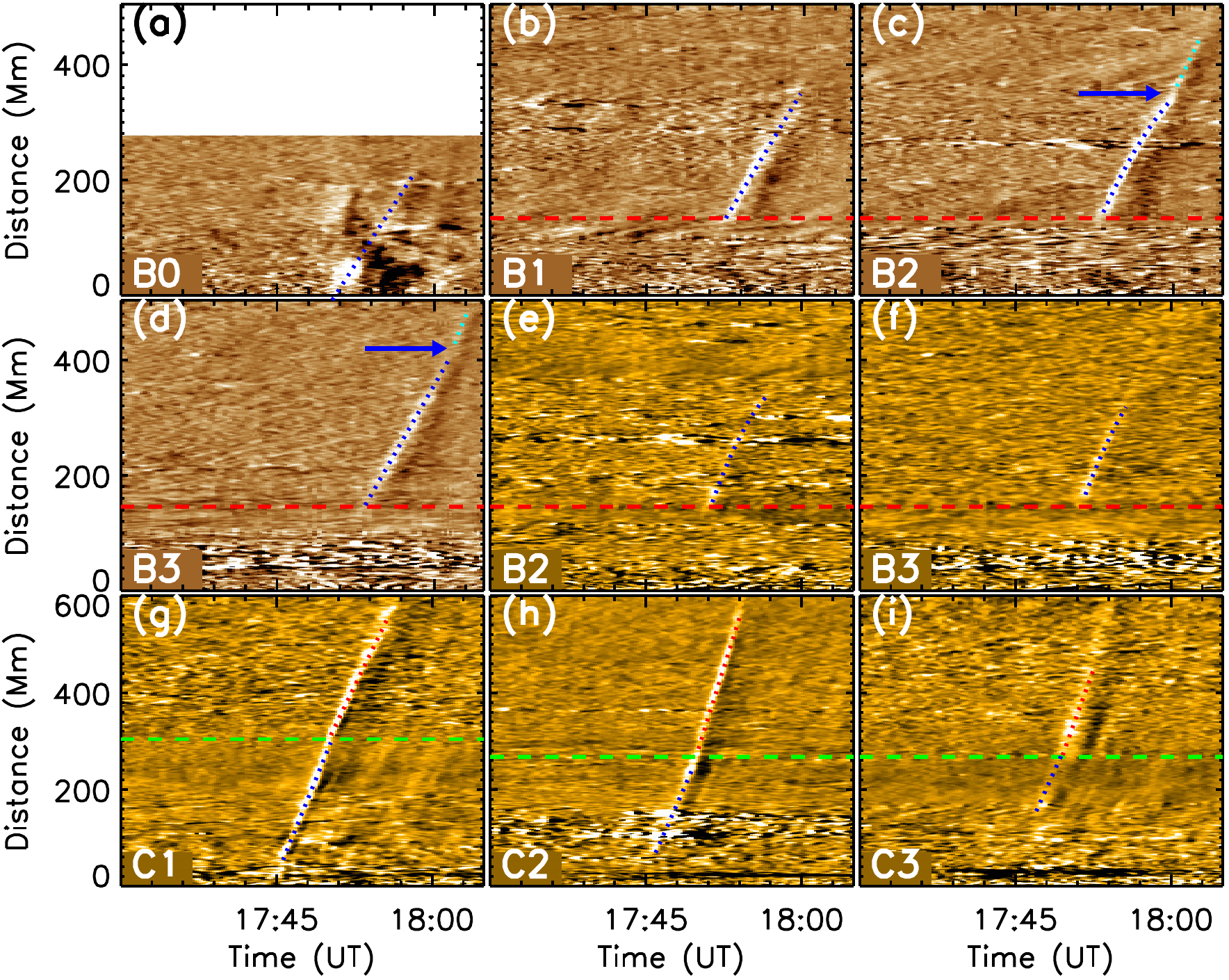}
\caption[]{\footnotesize
Time-distance plots along sectors B0 -- B3, and C1 -- C3. Panels (a) -- (d) and (e) -- (f) are obtained from 193 \AA\ and 171 \AA\ running ratio images respectively, while (g) -- (i) are obtained from 171 \AA\ running ratio images along C1 -- C3. The dashed (green) lines mark the outer boundary of AR11465 (AR11459). The dotted lines are linear or quadratic fit of the wave ridges. The blue arrows point to the refraction positions of the diffracted wave.
\label{fig4}}
\end{figure*}

We use time-distance plots to study the detailed kinematics of the primary EUV wave, as well as the reflected, diffracted, and refracted waves. The details of the method could be found in \cite{long11}. To minimize the spherical projection effect, three sets of great-circle sectors projected onto the Sun are used (see \fig{fig1}), in which sectors A1 -- A5 are used to measure the global behaviors of the primary wave, while the reflected, diffracted, and refracted waves are analyzed using sectors B0, B1 -- B3, and C1 -- C3, respectively. The identified origin of the waves acting as the crossing points of the great circles (see C1, C2 and C3 in \fig{fig1}), and the angle of each sector is set as 10${}^\circ$. From each image, a one-dimensional intensity profile as a function of distance within a sector can be obtained by averaging the intensities across the sector, in annuli of increasing radii with 0.05${}^\circ$ that corresponds a distance of 608 km. Composing the obtained profiles over time yields a two-dimensional time-distance plot, in which the propagating EUV wave shows up as an incline ridge, whose slope represents the apparent speed along the solar surface.

\fig{fig3} shows the time-distance plots made from 193 and 171 \AA\ running ratio images along sectors A1 -- A5. Along A1, the wave propagated into a region with open fields (see \fig{fig1} (a)), and a bright stationary front formed there (see the top row in \fig{fig3} and the red arrow). However, the wave kept its propagation rather than stopping there. This implies that the EUV wave penetrated through a topological separatrix surface, suggesting the wave property of this EUV wave. Sector A2 passes through the center of AR11465, the time-distance plot made along this sector well displays the interaction between the wave and AR11465 (second row in \fig{fig3}). It can be seen that the generation of the EUV wave was after the start of the loops' expansion, and the former was running ahead of the latter. This pattern suggests that the origin of the EUV wave was possibly driven by the expanding loops during the initial stage. As the wave approached AR11465, it became weaker and finally disappeared in the active region. However, about five minutes later, a new wavefront appeared on the other side of AR11465 (see the green arrows in \fig{fig3}), consistent with the image observational results. Sectors A3 and A4 are placed on the quiet-Sun region in-between AR11465 and AR11459. We can see that the EUV wave experienced a significant deceleration in the 193 \AA\ plot (see \fig{fig3} (c)), from which we obtain the linear wave speed (acceleration) is about \speed{493} (\acc{-416}), and the initial and final speed are \speed{806 and 350} respectively. This result suggests that the EUV wave is a typical nonlinear wave that gradually decays to an ordinary linear wave \citep{warm11,shen12a}. However, in the 171 \AA\ time-distance plot along sector A4, the wave ridge could be fitted with a straight line and that yields a speed of \speed{800}. Sector A5 passes through AR11459, and the time-distance plot along it well displays the transmission of the wave through the active region. We obtain that wave speeds before, during, and after the transmission are \speed{640, 804, and 1264}, respectively. It should be noted that the speeds during and after the transmission are not accurate due to the changing of the propagation direction, since the different properties of the mediums such as magnetic field strength and density.

To measure the wave speed after the interactions more accurate, we use two other sets of sectors as shown in \fig{fig1}, in which sectors B0, B1 -- B3, and C1 -- C3 are used to measure the speeds of the reflected, diffracted, and refracted waves, respectively. \fig{fig4} (a) shows the reflected wave in 193 \AA\ time-distance plot, it propagated with a speed of \speed{469}, close to that of the incoming wave (\speed{493}). By averaging the speeds of the diffracted wave measured along B1 -- B3, we obtain that the average speed of the diffracted wave in 193 \AA\ observations is about \speed{516}, while that is about \speed{617} in 171 \AA. It should be noted that refractions were also observed when the diffracted wave propagated about 260 Mm from the outer boundary of AR11465 (see the blue arrows in \fig{fig4}), where the wave suddenly changed its propagation direction. This is possibly caused by small magnetic structures such as coronal bright point as shown in \cite[e.g.,][]{shen12a}. The refraction of the EUV wave by AR11459 is shown in \fig{fig4} (g) -- (h). The average propagation speeds of the wave inside and outside of the active region are \speed{829} and \speed{956}, respectively. Along sectors C1 -- C3, the wave underwent different angles of refraction along these sectors at the outer boundary of AR11459, which depend on the incidence angle of the incoming wave.

\section{CONCLUSION AND DISCUSSIONS}
With high temporal and spatial resolution observations taken by {\em SDO}/AIA, we report the diffraction, refraction, and reflection of a global EUV wave during its interactions with two remote active regions in the other hemisphere. The diffraction and reflection were observed when the EUV wave interacted with AR11465, the average speeds of the primary, reflected, and diffracted waves are \speed{493, 469, and 516}, respectively. The primary EUV wave showed a significant deceleration and it could be considered as a typical nonlinear wave as defined by \cite{warm11}. The refraction of the diffracted wave was also observed, which is possibly caused by some small magnetic structures such as coronal bright points as reported by \cite{shen12a}. In addition, we find that the EUV wave can penetrate a magnetic region with open fields. The refraction was observed when the EUV wave transmitted through AR11459. Due to the different physical conditions inside and outside AR11459, during the transmission the wave changed its propagation direction at the both boundaries of the active region. The speeds before, during and after the transmission are \speed{640, 829, and 956}, respectively. All these observational results together exhibit the behavior of a true magnetohydrodynamics (MHD) wave. For the generation of the EUV wave, we find that it formed ahead of a group of expanding loops a few minutes after the start of the loops' expansion, which may represent the initiation of the associated CME. These results indicate that the launching of the EUV wave should be driven by the associated CME. Based on the observational results, we conclude that the EUV wave should be a nonlinear magnetosonic wave driven by the associated CME during the initial stage. 

\begin{figure*}[thbp]
\epsscale{0.9}
\plotone{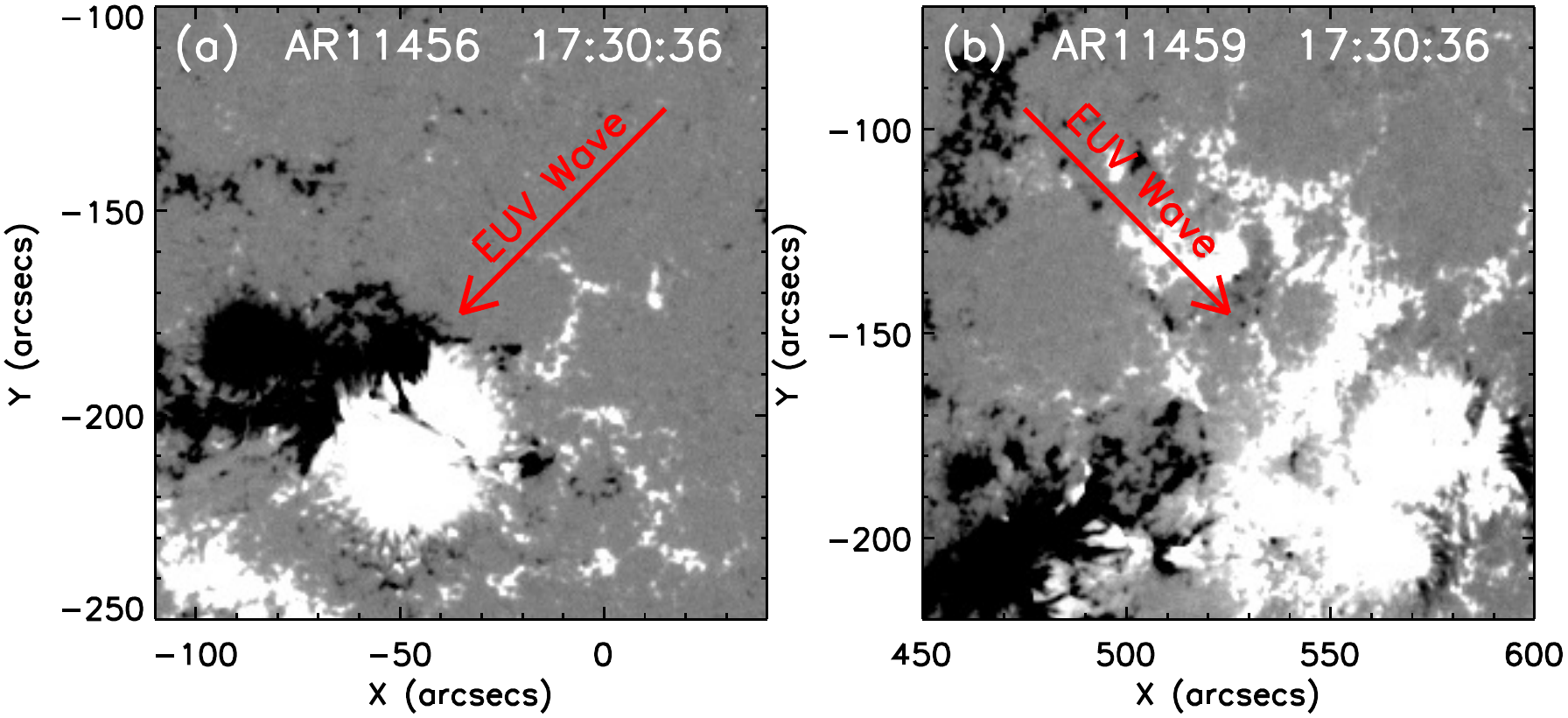}
\caption[]{\footnotesize
The pre-event HMI LOS magnetogram at 17:30:36 UT show the magnetic configurations of AR11465 (a) and AR11459 (b), in which white (black) color represents the positive (negative) polarity. Their field-of-views are indicate by the pink boxes in \fig{fig1}(a). The red arrows indicate the direction of the incoming EUV wave.
\label{fig5}}
\end{figure*}

The EUV wave presented here could be interpreted with the so-called hybrid model \citep[e.g.,][]{chen02,chen05,cohe09,down11}, in which a fast magnetosonic or shock wave travels ahead of a slow compression front of the surrounding medium caused by the expansion of the erupting structures. In the present cace, the EUV wave corresponds to the fast component in the hybrid model, while the expanding loops represents the slow apparent wave component. It should be noted that low cadence observations in the past could lead to different interpretations on EUV waves. For example, low temporal resolution observations may miss the leading fast wave component in some cases and only capture the trailing non-wave front caused by magnetic reconfiguration \citep[e.g.,][]{liu10, liu12}. In addition, many studies have indicated that EUV waves have a nature of deceleration, especially during the initial stage, that would result in low derived speeds for observations with a low temporal cadence \citep[e.g.,][]{warm01,warm11,vrsn06,vero08,shen12b}.

For the appearance of the new wavefront on the southeast of AR11465, we explain it as the diffracted wave caused by the active region. Due to the strong magnetic field strength in the active region, it may exist a large speed gradient at its boundary \citep{schm10,liu12}. If the speed gradient is large enough, the incoming wave can not penetrate the active region. On this occasion, one can expect simultaneously a diffracted wave behind the active region and a reflected wave on the wave incoming side, as indicated by our observation. In the same line of thought, the refraction observed in AR11459 could be interpreted as the strong transmission of the EUV wave due to a small speed gradient at the  boundary of AR11459. Due to the lack of vector magnetic field of the two active regions, we can not obtain the accurate coronal field above with some sophisticated extrapolation methods. Therefore, we just show the photosphere line-of-sight (LOS) magnetogram in \fig{fig5} to exhibit the magnetic configurations of the two active regions, which may demonstrate the above coronal magnetic field to some extent. On the side of the incoming wave, the magnetic field distribution of AR11459 is more diffuse than that of AR11465. In addition, the maximum value of the magnetic field strength in AR11459 (1355 Gauss) is smaller than that of AR11465 (1924 Gauss). These result may suggest a larger speed gradient at the boundary of AR11465 than that of AR11459, that is possibly why the EUV wave was diffracted (refracted) at AR11465 (AR11459). The critical value of speed gradient for the occurrence of reflection, diffraction and refraction of the a EUV wave should be important for us to understand the coronal physics. However, we can not estimate it based on the current single event. We think that the numerical simulation could be a helpful solution for this problem. On the other hand, due to the strong magnetic field strength in the active regions, the wave speed in the active region should be much higher than that in the quiet-Sun, and the compression of plasma should be lower in the active region. Taking these factors and the circular shape of the active region into consideration, one can also expect the circular secondary wave behind the active region and the disappearance of the EUV wave in the active region.

The diffraction of EUV waves is probably a common phenomenon in the solar corona. For example, in the wave event on 2011 February 15, which had been studied by \cite{schr11,olme12}, we note that diffraction of the wave was occurred when the wave interacts with the active region located in the northern hemisphere, but the authors did not analyze this interesting phenomenon (see animations 1 and 2 in \cite{olme12}). To fully understand the diffraction of global EUV waves by coronal magnetic structures, more similar case studies as well as theoretical analysis are desirable.

\acknowledgments We thank the observations provided by the {\em SDO}, which is a mission for NASA's Living With a Star (LWS) Program. The authors would like to thank an anonymous referee for many valuable suggestions for improving the quality of this paper. We also thank Dr. S. Liu, and Mr. Y. Bi for their help. This work is supported by the NSFC (1093303, 11078004, 11073050), the 973 Program (2011CB811400), the CAS Programs (KJCX2-EW-T07 and 2012Y1JA0002), the Open Research Programs of the Key Laboratory of Solar Activity (KLSA201204, KLSA201219) and the Key Laboratory of Dark Matter and Space Astronomy of CAS (DMS2012KT008), and the Western Light Youth Project.

\end{document}